\documentclass[11pt]{article}
\usepackage{graphicx}

\title{\bf Quantization of energy and writhe in self-repelling knots}
  
\author{{Phoebe Hoidn$^{1}$,  Robert B. Kusner$^{2}$ and
 Andrzej Stasiak$^{1}$} \\ \\
\em  $^{1}$ Laboratoire d'Analyse  Ultrastructurale, Universit\' e 
de Lausanne,\\
\em 1015 Lausanne, Switzerland.\\   E-mails: Phoebe.Hoidn@lau.unil.ch, Andrzej.Stasiak@lau.unil.ch\\  \\
\em  $^{2}$ Department of Mathematics, University of Massachusetts, 
Amherst,\\
\em MA 01003-4515, USA. E-mail: kusner@math.umass.edu } 
\date{}

\begin{document}
\maketitle

\begin{abstract}
Probably the most natural energy functional to be considered for knotted strings is the one 
given by the electrostatic repulsion. In the absence of counter-charges, a charged, knotted
 string evolving along the energy gradient of electrostatic repulsion would progressively tighten 
 its knotted domain into a point on a perfectly circular string. However, in the presence of charge
  screening the self-repelling knotted strings can get stabilised. It is known that energy functionals 
  in which repulsive forces between repelling charges grow inversely proportional to the third or
   higher power of their relative distance stabilise self-repelling knots. Especially interesting 
   is the case of the third power since the repulsive energy becomes scale invariant and does not 
   change upon  M\"{o}bius transformations (reflections in spheres) of knotted trajectories. We observe 
   here that knots minimising their repulsive   M\"{o}bius energy show quantization of the energy and writhe 
   (measure of chirality) within several tested families of knots.
   \vspace{10mm}
\end{abstract}

\section{Introduction} \vspace{10 mm}

Knot theory was given a strong impetus when in 1860's Kelvin proposed that knots made 
out of vortex lines of ether constitute elementary particles of matter which at this time 
were thought to be atoms \cite{1}. However, development of atomic physics, first with classical
and then with quantum models of atoms, failed to show a connection between knots and atoms.
More recently though, today's elementary particles are again considered to be string like 
objects that may be closed \cite{2, 3} and perhaps knotted. This string theory approach partially
revives the ideas of Kelvin and provides a motivation for exploring quantization of energy
and of other values in such physical systems such as knotted magnetic flux lines \cite{4, 5} or
knotted solitons \cite{6}.\\

Knots made of self-repelling, infinitely thin strings in 3-dimensional Euclidean space have 
been considered by many authors \cite{7,8,9,10,11,12}. Usually it was assumed that knots were perfectly 
flexible for bending but not extensible. The repelling charge was assumed to be continually spread along 
the knots so that there were no elementary point charges and that the charge contained within a 
given curvilinear segment is proportional to the length of this segment. When the Coulomb energy
would govern the evolution of a knot of this type one should observe that the knotted domain gets 
progressively tightened into a singular point while the rest of the knot would form a perfect
circle. Progressive tightening of knotted domains in electrostatic knots may seem 
counterintuitive when one considers that the electrostatic repulsion grows inversely
proportional to the square of the distance between every pair of charges. However, 
   at the same time as tightening progresses, the length of the knotted domain decreases
    and therefore there is less and less charge in the tightly knotted domain. 
    In fact the
     decrease of charge in the knotted domain is more rapid than its contribution 
     to overall 
     repulsion. Therefore the tightening that is driven by the decrease of the repulsion 
     outside of the knotted domain can progress until a knotted domain shrinks to a singular 
     point and the entire string is  perfectly circular.
      For this reason the Coulomb energy is not interesting as energy functional for 
     knots \cite{13, 14}. However, it had been demonstrated that if the repulsion force
      were growing inversely proportional to the third or higher power of the distance 
      between the repelling elements then the knotted domains in prime knots would have no 
      tendency to shrink to a singular point \cite{7, 8, 10, 12}. The third power case is especially 
      interesting from mathematical and physical points of view since the energy of a knot becomes
       conformal and therefore does not change when the trajectory of the knot is rescaled or
        undergoes a M\"{o}bius transformation (reflection in a sphere) \cite{10, 14, 15}.  In this work 
	we study configurations of knots that minimise their M\"{o}bius repulsive energy. From now 
	on we will call these configurations as M\"{o}bius knots. Several earlier studies used 
	numerical simulations or analytical approach to investigate various properties of
	 M\"{o}bius knots \cite{9, 15}, however the relations between such characteristic properties 
	 of M\"{o}bius knots as their energy, crossing number, writhe or average crossing number 
	 were not systematically examined before.

\vspace{2cm}

\section{Energy of M\"{o}bius knots}\vspace{10mm}

Let us consider an unknotted closed string that has a few repulsive point charges equidistantly
 separated. An energy minimising shape of such an unknotted string would be then an equilateral
  polygon with the number of vertices corresponding to the number of the point charges. To be 
  able to operate with a model of self-repulsive knot whose shape is independent of the number
   of charges in the knot one needs to assume that charges are not localised but continuously 
   spread over the knot. This mathematical operation ensures that unknotted energy-minimising 
   strings would always form a perfect circle independently of the level of carried charge. 
   However, this non-physical assumption of continuous charge redistribution causes the energy 
   of a knot to become infinite due to the repulsion of nearby elements. In order to correct 
   for this problem of infinite energies O'Hara \cite{8} introduced a regularisation term and defined
    the energy
\begin{equation}
\label{one}
\widetilde E(K) = \int\!\!\!\int_{K \times K} \left( {1 \over {\vert x - y \vert ^2}} 
- {1 \over {d_K(x,y)^2}} \right) 
ds_K (x) ds_K (y) 
\end{equation}   
where $d_K(x,y)$ is the shorter arclength distance within $K$ from $x$ to $y$. Notice that the integral 
of the second term corresponds to the repulsive energy of a straight segment with the same length 
and carrying the same charge as $K$. Another, computationally more stable approach is to neglect
 tangential contributions to repulsion as nearest neighbour regions in smooth trajectories are 
 practically co-linear and define the cosine energy
 
 \begin{equation}
\label{two}
E (K) = \int\!\!\!\int_{K \times K}  {{ (1 - \cos \alpha) }\over 
{\vert x - y \vert ^2}}  
ds_K (x) ds_K (y) 
\end{equation} 
where $\alpha$ is the conformal angle between the tangents at points $x$ and $y$. In fact it was
 demonstrated by Doyle and Schramm \cite{15} that 
 \begin{equation}
\label{three}
\widetilde E(K) = E(K) - 4 .
 \end{equation} 
 
 We have applied here the second approach and used Kenneth Brakke's program
 Evolver to obtain 
M\"{o}bius energy minimising configurations of various knots and to 
calculate their energies 
(Ref. 16 but see also http://www.\break susqu.edu/facstaff/b/brakke/). 
Examples of configurations
 of various knots minimising their M\"{o}bius energy can be seen in Reference 15. It should be
  recalled that actual shapes of M\"{o}bius energy minimizers of a given knot can substantially
   vary because all configurations obtained by M\"{o}bius transformations from one  M\"{o}bius energy 
   minimizer are also M\"{o}bius energy minimizers. In practice these shapes depend on an arbitrary 
   choice of starting configurations used for the energy descent. However, the actual energy values
    obtained in our simulations converge to the same values independently of the starting
     configurations of a given knot. In addition we have checked that the obtained by us values 
     for ten different $(2,p)$ torus knots were at most different by 0.1 \% from the values 
     calculated using an analytical approach that can be applied to this class of knots \cite{9, 15}. 
 \begin {figure} 
 \hspace{-1cm}
\includegraphics[width= 14cm]{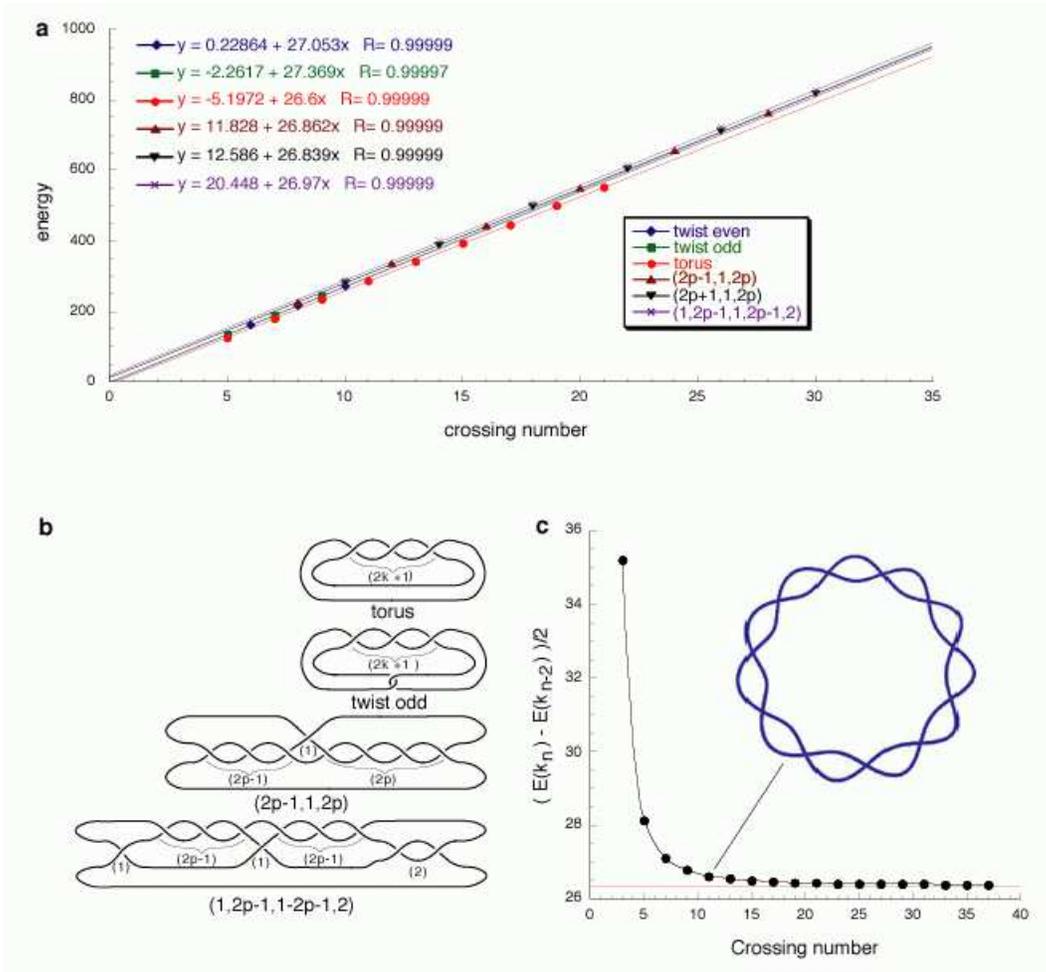}
\\

\caption{\em  Energy of M\"{o}bius knots, {\bf a}, Different families of torus knots, twist knots and Conway
knots show the same slope. These analysed families are represented in {\bf b} with $k = 1$ and $p = 2$. {\bf c}, The 
difference of energy between sequential torus knots tends asymptotically towards 52.8 which corresponds to
26.4 for each new 
crossing. The inset shows one of the configurations of the torus knot $11_1$ that minimises the 
M\"{o}bius repulsive energy, its position on the plot is indicated.}  
\label{Fig.1}
\end{figure}      
    
Figure 1 shows the relation between the M\"{o}bius energy and the topological, minimal crossing number
 for knots belonging to six different families of knots. We have analyzed torus knots with
  Alexander-Briggs notation $3_1, 5_1, 7_1$ etc, twist knots with even number of crossings 
  (Alexander-Briggs notation $4_1, 6_1, 8_1$ etc), twist knots with odd number of crossings
   (Alexander-Briggs notation $3_1, 5_2, 7_2, 9_2$ etc) and three Conway families of knots with 
   Conway notations $(2p+1,1,2p), (2p-1,2,2p)$ and $(1,2p-1,1,2p-1)$ where $p$ are consecutive natural
    numbers (see Fig. 1b for schematic explanation how these families are formed). Standard 
    representations of these knots classified according to Alexander-Briggs notations can be 
    seen in tables of knots \cite{17, 18} while the M\"{o}bius energy minimising configurations are shown 
    in Ref. 15 and one representative example of energy minimising configurations of torus knot 
    $11_1$ is shown here in Figure 1c. From the data points in Figure 1a we have excluded founders 
    of the families as these frequently belong to different families at the same time. It
     is visible that in all these knot families the energy grows with a practically identical 
     rate that seems to be linear. Linear fit over the tested range indicates that for each new 
     crossing the energy grows by around 26-27 units. To give an estimation of energy units it
      is good to point out here that within the energy defined by the equation (\ref{one})  the energetic 
      costs of closing an open string into unknotted circle is exactly 4 units \cite{14}.
       Notice that
       some of the plotted lines in Figure 1a practically coincide with each other while 
       other 
       seem to be vertically shifted by a constant value.\\

 A closer look at energy values for the $(2,p)$ torus knot family $(3_1, 5_1, 7_1$ etc.), including 
 the founder $3_1$ knot, suggests that the energy difference between consecutive torus knots shows 
 an asymptotic convergence toward a constant value of circa 53. Since the energy minimisation by
  simulation with Evolver (or other programs) has its well known limitations in descending toward 
  a global minimum, especially for complex trajectories, we have used an analytical approach to
   generate torus knots configurations that are believed to be minimizers of the M\"{o}bius energy 
   \cite{9, 15}. Figure 1c shows that the difference of energy between sequential torus knots 
   tends asymptotically toward 52.8 which corresponds to 26.4 for each new crossing. \\
   
On the basis of our simulation data (Fig. 1a) and analytical approach (Fig. 1c) we conjecture 
that within such families of knots that iteratively increase their interwound regions with
 double helix structure \cite{15}, the differences in M\"{o}bius energy due to each new crossing tend 
 to an universal constant value.
 
 \vspace{2cm}

\section{Writhe of M\"{o}bius knots} \vspace{10mm}

As already discussed, knots minimising M\"{o}bius energy do not have unique shapes since 
M\"{o}bius transformations can create infinitely many different configurations that minimise
 the M\"{o}bius energy for a given knot. However, certain characteristic properties of curves 
 in space are not changed by M\"{o}bius transformation. It was proven by Banchoff and White that the 
 absolute value of writhe for a given trajectory is invariant upon M\"{o}bius
  transformation \cite{19}.
  Writhe (Wr) measures the extent of chirality of closed curves in space and therefore provides an 
  interesting measure for knotted trajectories minimising a given energy functional.
   The writhe
   corresponds to the average difference between numbers of right- and left-handed crossings
    perceived when a given curve in space is observed from a random direction.
     The value of 
    writhe (Wr) is usually calculated using the Gauss integral formula
    \begin{equation}
\label{four}
\hbox{Wr} = \int\!\!\!\int_{K \times K} {{(u \times v) \cdot (x - y)} 
\over 
{\vert x - y \vert ^3}}  
ds_K (x) ds_K (y) 
\end{equation}  
where $u$ and $v$ are the unit tangent vectors to $K$ at $x$ and $y$, respectively. In the
 case of tight 
knots minimising their rope length and that are known as ideal, the writhe values showed a
 quantization \cite{20,21,22} and the quantum of writhe depended on the type of introduced
  crossings 
 \cite{23, 24}. We decided therefore to check whether knots minimising their M\"{o}bius energy
  also show a similar quantization of writhe.
 
  \begin{figure} 
  \hspace{-1cm}
\includegraphics[width = 14.5 cm]{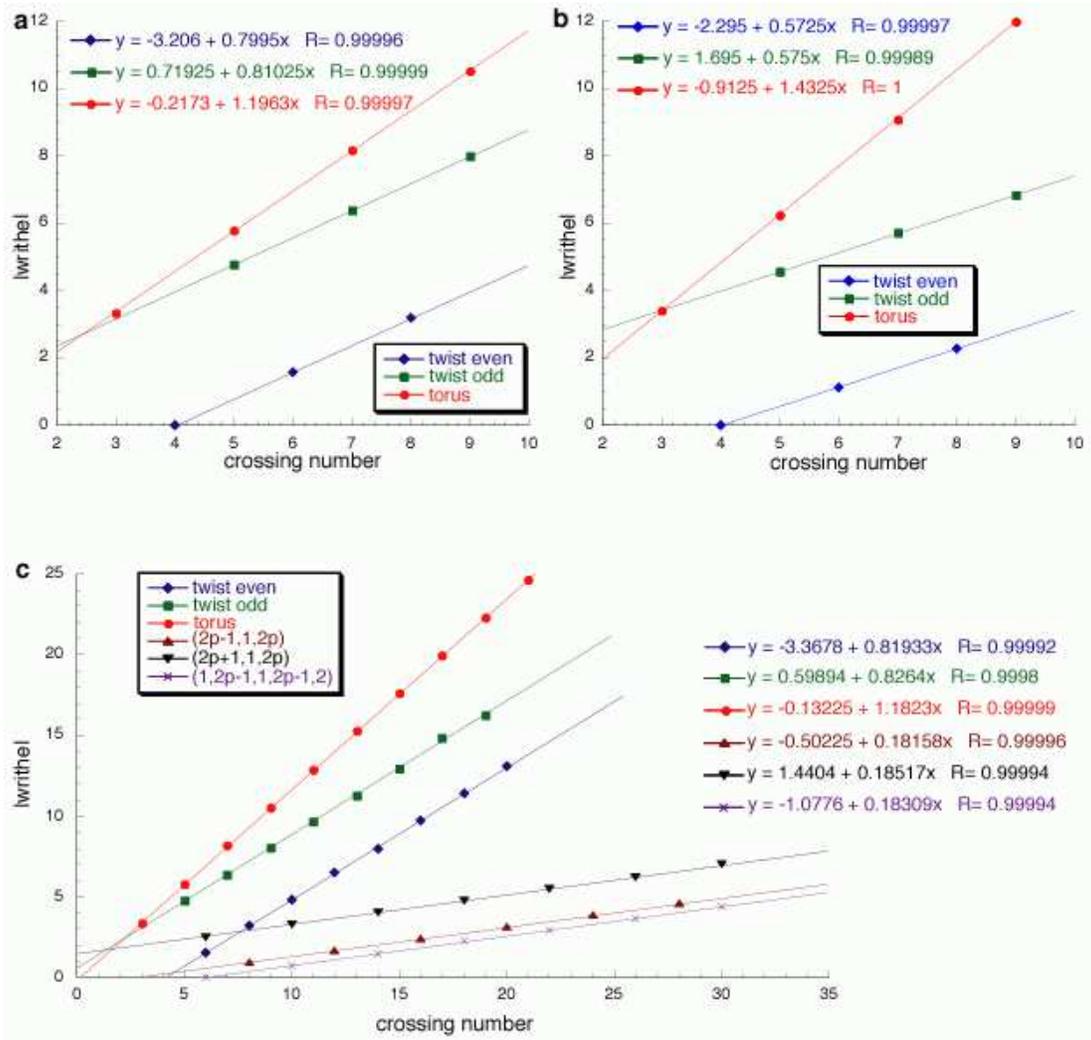} 
\\

\caption{\em Writhe of M\"{o}bius knots. {\bf a} and {\bf b} show the quantization of absolute
writhe of M\"{o}bius knots and ideal knots, respectively, the analysed families are torus knots
and twist knots. {\bf c}, Absolute writhe slopes for all tested by us families of knots.}  
\label{Fig.2}

\end{figure}

Figure 2a illustrates that M\"{o}bius torus and twist families of knots show a clear quantization
 of writhe where the writhe increase for torus knots is more rapid than for twist knots. While
  the differences of energy within a given family were showing an asymptotic descent toward a 
  limiting constant value (Fig. 1c), the differences of writhe showed a specific constant value
   (within the accuracy of our computational approach) that seemed to be independent of the
    complexity of the knot. In case of ideal knots (see Fig. 2b) it was observed earlier that
     the slopes of writhe increase for torus knots and twist knots were described by simple
      relations: $1 + x$ and $1- x$ respectively \cite{23, 24}, where $x$ was the same for both
       families
       and seemed to correspond to 3/7 \cite{23, 24} and where the integer values $1$ 
       (or $-1$ depending on the handedness of crossing) is due to inter-coil crossing contribution 
       while the noninteger value $x$ is due to intra-coil contribution to writhe \cite{23}. 
       We observed here that the slopes of writhe increase for M\"{o}bius torus and twist knots
        also show opposing deflections from the slope of one, however the deflection value $x$ is
	 close to 0.2 instead of 0.43. Thus torus-type of turns with positive signs of crossings 
	 introduce a writhe of circa 1.2 per crossing. Twist-type of turns with positive signs of
	  crossings introduce a writhe of circa 0.8 per crossing. Torus- and twist-type of turns 
	  can be recognised by the orientation along both arcs enclosing double-arc fields in
	   minimal crossing diagrams of a knot \cite{23}. Parallel and anti-parallel orientations 
	   characterise torus- and twist-type of turns, respectively \cite{23}.\\
	   
In the case of ideal knots it was observed that the writhe of achiral knots was essentially 
equal to zero \cite{20, 24}. We observed here the same tendency for knots minimising their M\"{o}bius
 energy as all achiral knots tested by us (i.e. $4_1, 6_3, 8_3, 8_9, 8_{12}, 8_{17}$ and $8_{18}$) had their 
 writhe practically equal to zero.\\
 
Figure 2c shows absolute writhe slopes for all families of knots tested by us. It is visible 
that within each of these families of knots there is apparently constant, specific increase of
 writhe as one analyses consecutive members of respective families. Interestingly the increase
  of writhe can be simply predicted by analysing the type of turns that are introduced while 
  creating a next member of the family. So for example in the $(2p+1,1,2p)$ Conway family, 
  as one goes from a knot to its successor four new crossings are introduced: two are positive 
  torus crossings which increase the writhe by about 2.4 and two are negative twist crossing 
  which decrease the writhe by about 1.6. Therefore the predicted increase of writhe of about
   0.8 when divided by four crossings gives us the observed slope about  0.2 (see Fig. 2b).
    We similarly can predict and explain why knots from even and odd twist families of knots 
    follow slopes with the same inclination but that are vertically shifted by exactly four
     units in relation to each other. Figure 3 
  \begin {figure} 
  \hspace{0.3cm}
\includegraphics[width= 12cm]{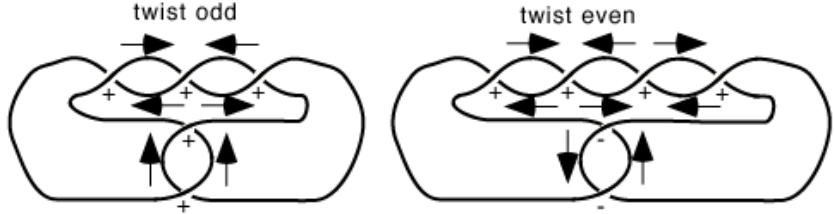}
\caption{\em   Odd and even twist
knots differ in the sign and type of crossing in the terminal clasp. Odd twist knots have terminal clasp
with torus type of crossings orientations along the arcs enclosing this double-arc field is parallel) 
while
even twist knots have twist type of 
crossings (orientations along the arcs enclosing this double-arc field is anti-parallel) Notice that
 sign of
crossings in the terminal clasp also changes although signs and types of crossings in the inter-wound region
remain the same.}  
\label{Fig.2}
\end{figure}         
     shows that change from an odd to an even twist
      family of knots implicates that two torus type of crossings in the terminal clasp are 
      replaced by two twist type of crossings of opposite sign. Since the corresponding writhe
       contributions of these crossings are $1+x$ and $-1+x$ (or $-1-x$ and $1-x$ for opposite 
       handedness), where $x$ is a constant, the resulting absolute difference of writhe is
        two per crossing irrespectively of the actual value of intra-turn contribution $(x)$ to 
	writhe. Since there is a change of two crossings the global difference of writhe between 
	twist even and twist odd families of knots results in the observed vertical shift of 
	corresponding writhe slopes by exactly four. We can apply a very similar type of 
	reasoning to explain why Conway families $(2p+1,1,2p)$ and $(2p-1,1,2p)$ differ in their
	 corresponding values of writhe by exactly 2. Notice that the same type of
	  explanation (where intra-turn contributions to writhe cancel) applies not only
	   to M\"{o}bius knots but also to ideal knots. Figure 2b shows that in the case of 
	   ideal knots, odd and even twist knots also show the relative vertical shift 
	   by exactly four although the actual slope values are different. Axial trajectories 
	   of ideal knots (rope length minimizers) can be regarded as a limit of energy minimising 
	   configurations when the energy is taken to be an integral of an ever-increasing 
	   inverse-power of the radius of certain circles passing through three points of the
	    curve \cite{25}. Therefore this observation strongly suggest that minimizers of
	     repulsive functionals with any exponent between 3 and ƒ$\infty$ will always have a constant
	      shift of writhe slopes between certain families of knots. Therefore for example 
	      slopes of absolute writhe for even and odd twist knots will always show a relative 
	      vertical shift by exactly four (compare Fig. 2a and b) provided that 
	      compaired knots minimize the same repulsive functional 
	      with the exponent ranging between 3 and $\infty$.    
 
 \vspace{2cm}

\section{Relations between energy and crossings} \vspace{10mm}	      
 
There are two principal measures of crossings applied to knots.
 Minimal crossing number is a topological invariant and corresponds to 
 the minimal number of crossings that any representation of this knot type can have
  in any orthogonal projection.  Average crossing number (ACN) applies to a given rigid 
  embedding of a knot and corresponds to the average number of perceived crossings 
  (irrespective of their handedness) when this particular embedding is perceived from a random
   point on a sphere enclosing a given trajectory. The ACN value can be calculated using the 
   unsigned Gauss integral formula 
      \begin{equation}
\label{five}
\hbox{ACN} = \int\!\!\!\int_{K \times K} {{\vert(u \times v) \cdot (x - y)\vert} 
\over 
{\vert x - y \vert ^3}}  
ds_K (x) ds_K (y) 
\end{equation}  
where $u$ and $v$ are the unit tangent vectors to $K$ at $x$ and $y$, respectively. 
Several studies considered theoretically the relation between the M\"{o}bius invariant repulsive 
energy and the two measures of crossings mentioned above. Freedman, He and Wang demonstrated 
for example that the M\"{o}bius energy of a knot is at least $2  \pi$ fold bigger than the minimal 
crossing number of this knot \cite{10}. 
  \begin {figure} 
  \hspace{-1cm}
\includegraphics[width= 14.5cm]{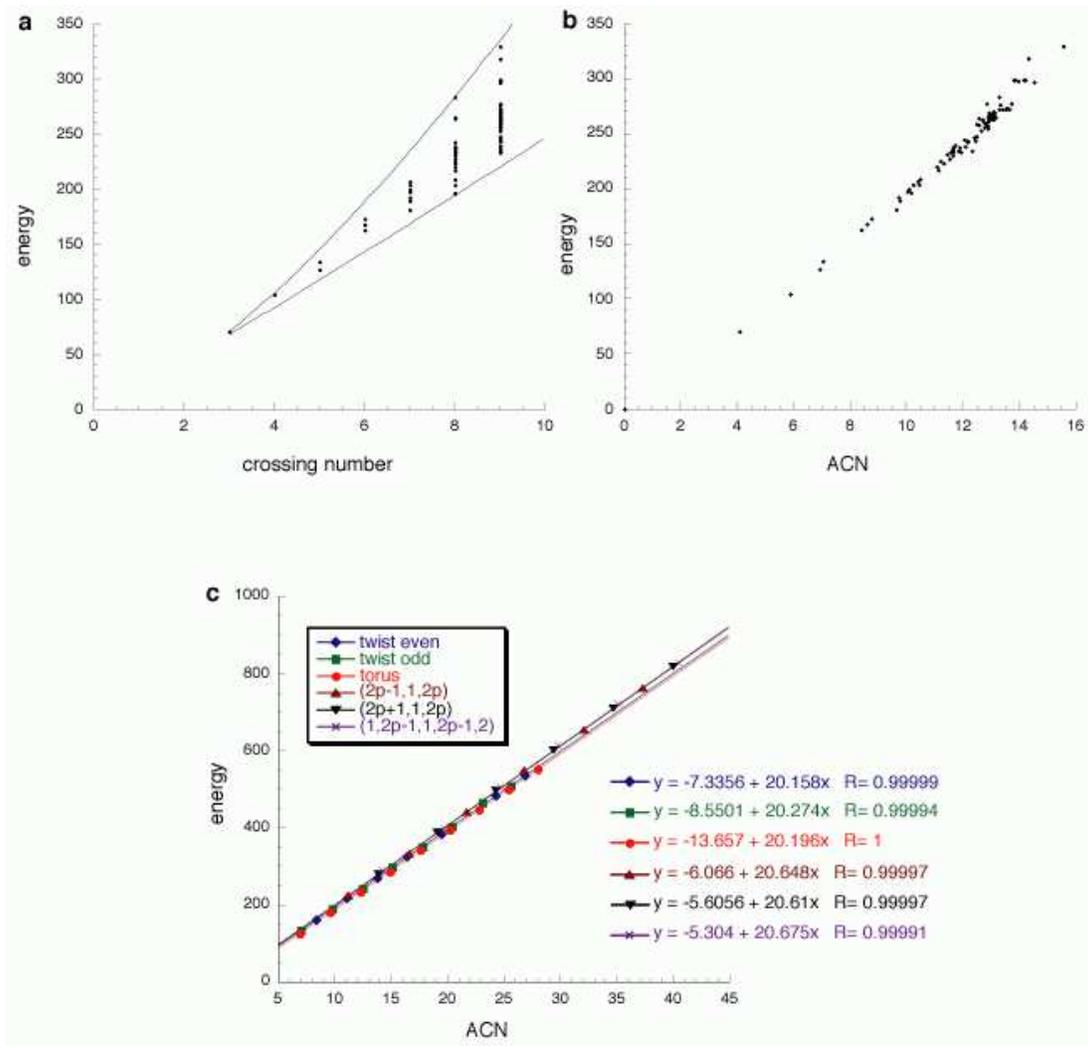}
\\

\caption{\em Relation between the energy and crossings for all prime knots 
up to 9 crossings. {\bf a},
The relation between M\"{o}bius energy and minimal crossing number for all
 prime knots up
to 9 crossings. The upper on lower bounds were best fits of power law
 functions $y = ax^n$. Fits satisfied
the 
condition that no experimental point was outside of 
the bounds. {\bf b}, The relation between M\"{o}bius energy and ACN for all prime knots up to 9 crossings. 
{\bf c}, The
energy as a function of average crossing number within different families of knots.}  
\label{Fig.3}
\end{figure}
Figure 4a shows the relation between M\"{o}bius energy 
and minimal crossing number of all knots up to 9 crossings. It is visible that data points 
fill a  ``cone'' and that for these relatively simple prime knots the lower linear bound of the  
energy could be put at least at $7.47 \pi  $ times the minimal crossing number. One can also analyse 
the upper bound for the energy as the function of minimal crossing number. In figure 1a we have
 shown that within a given family the energy grows proportional to the crossing number, but as 
 the crossing number increases, founders of new families arise and these can have very high energy
  as compared to the members of already established families. Thus for example a twist knot
   $8_1$ has the energy of 217.4 while the $8_{18}$ knot founding a new family has the energy of 283.9. 
   The upper bound of the energy does not follow a linear relation with the 
   crossing number but
    approaches a 7/5 power law (see Fig. 4a). The ACN value is not invariant upon M\"{o}bius 
    transformation, however, we have checked that ACN value is very robust and hardly changes
     upon multiple M\"{o}bius transformations. Therefore it seems reasonable to investigate the
      relation between the energy and ACN of M\"{o}bius, knots.  Freedman, He and Wang showed that 
      the M\"{o}bius energy of a knot increased by 56/11 is at least $12  \pi /11$ fold times bigger than 
      ACN of energy minimising configurations \cite{10}. Existence of this linear
       lower bound
       demonstrates that the M\"{o}bius energy can not grow with a power lower 
       than 1 as a function 
       of ACN. It could though grow with the higher power. In fact our own data
        (Fig. 4b) indicate
        that data points for the relation between the energy and ACN fill 
	much narrower  ``cone''
	 than it was the case for the relation between the energy and minimal
	  crossing number 
	 (see Fig. 4a). The lower bound of this cone can be described by a nearly 
	 linear 
	 function, this function is given by 
	    \begin{equation}
\label{six}
 y = 22.25 \cdot x ^ {1.05}
\end{equation}  
whereas the upper bound is better described by a nearly 7/5 power law, the function is here  
   \begin{equation}
\label{seven}
 y = 14.78 \cdot x^ {1.42}
\end{equation}  
This power law behaviour may seem inconsistent with a linear growth of energy with the ACN
 within different analysed families of knots (Fig. 4c). However as the ACN increases new founders 
 of knots' families enter and they frequently start their families with a linear growth but
  at a higher level. Again $8_1$ and $8_{18}$ knots are good examples as the difference of their 
  energy
   is much bigger than expected from a slightly higher ACN of $8_{18}$ knot.
   
\vspace{2cm}

\section{M\"{o}bius knots and their relation to random knots and other knotted physical systems}
\vspace{10mm}
It was shown earlier that the average writhe for a population of random knots of a given type
 closely corresponds to the writhe of ideal knot of the corresponding type \cite{20, 26}. 
 We have shown here that with the exception of achiral knots the writhe values of M\"{o}bius knots
  do not correspond to that of ideal knots. Although within given families there is a linear
   relation between writhe of ideal and M\"{o}bius knots these relations are not universal and 
   different families will be related by different linear relations. Thus for example the
    writhe of torus knots grows more quickly for ideal knots than for M\"{o}bius knots but the
     opposite is true for twist knots. Therefore we can conclude here that the
      writhe of 
     M\"{o}bius knots is not related to the writhe of random knots of corresponding 
     type.\\
     
Time averaged writhe value of randomly fluctuating knots of a given type seems to be independent
 of the length of random chains forming a given knot. This is not the case of ACN 
 (average crossing number) as its value progressively increases with the length of a random chain.
  However, it was observed that for relatively simple knots the differences between time 
  averaged ACNs of randomly fluctuating knots of a given type and of unknots of the same 
  chain size closely corresponds to the ACN of ideal knots of the corresponding types \cite{14}.
   However this does not apply to ACN values of M\"{o}bius knots since these are significantly 
   smaller than ACN values of corresponding ideal knots. We conclude therefore that M\"{o}bius
    knots in contrast to ideal knots are not good predictors of certain physical properties of 
    random knots of a given type such as knotted polymer molecules undergoing a random thermal
     motion. However for other physical knotted systems, M\"{o}bius knots may better approximate 
     their behaviour than ideal knots. If we imagine a charged knotted string of dimensions 
     comparable to an effective screening radius at given conditions then all pair-wise
      interactions within such a knot should be repulsive. However, due to screening interactions 
      caused by counterions the repulsion would not follow the Coulomb law but would decrease 
      more rapidly with the separating distance, approaching perhaps a cubic root dependence 
      of the distance. Short flexible polymeric molecules like single-stranded DNA can make 
      knots of dimensions comparable to effective screening radius at specific ionic conditions, 
      such knots may approach then the M\"{o}bius behaviour. Finally one can entertain a thought about
       string-like charged elementary particles (electrons, for example) surrounded by short-lived
        mixture of other charged particles and antiparticles generated from quantum fluctuation
	 of vacuum. Electrons may minimise then an energy that resembles the M\"{o}bius energy 
	 described here. If on the way to relaxation a complex self-repelling knot could 
	 undergo from time to time a strand passage and progressively simplify its type
	  that would provide a physical system with natural quantization of the energy.

\vspace{1.5mm}

 {\bf Acknowledgments.} We thank Fran\c cois Ubertini and Akos Dobay for their help in 
solving frequent software and hardware problems, Piotr Pieranski and Corinne Cerf for
 discussions on writhe quantization, Jun O'Hara and John Maddocks for discussions on energy of 
 knots, Jacques Dubochet for his keen interest and constant encouragement. This work was 
 supported by Swiss National Science Foundation Grant 31-61636.00 to A.S..

\end{document}